\documentclass[journal]{IEEEtran}
\usepackage{color}
\usepackage{multirow}
\usepackage{cite}
\usepackage{balance}
\usepackage{amsmath}
\usepackage{amssymb}
\usepackage{enumitem}
\usepackage{dblfloatfix}
\setlist[itemize]{noitemsep, topsep=0pt}
\usepackage{amssymb}
\usepackage[utf8]{inputenc}
\usepackage[english]{babel}
\usepackage[usenames, dvipsnames]{xcolor}
\usepackage{pdfsync}
\usepackage{amssymb}
\DeclareMathAlphabet{\mathpzc}{OT1}{pzc}{m}{it}
\setlist[itemize]{noitemsep, topsep=0pt}
\usepackage{pdfsync}

\ifCLASSINFOpdf
\usepackage[pdftex]{graphicx}
\DeclareGraphicsExtensions{.jpg}
\else
\usepackage[dvips]{graphicx}
\DeclareGraphicsExtensions{.eps}
\fi
\graphicspath{{figs/}}
\usepackage{epstopdf}
\usepackage{booktabs}
\usepackage{amsmath}
\usepackage{soul,xcolor}

\begin{document}

\title{Joint Estimation of Operational Topology and Outages for Unbalanced Power Distribution Systems}

\author{Anandini Gandluru,~\IEEEmembership{Student Member,~IEEE}, Shiva Poudel,~\IEEEmembership{Student Member,~IEEE}, and Anamika Dubey,~\IEEEmembership{Member,~IEEE}
	\thanks{A. Gandluru, S. Poudel, and A. Dubey  are with the School of Electrical Engineering and Computer Science, Washington State University, Pullman, WA, 99164 e-mail: a.gandlurunaga@wsu.edu, shiva.poudel@wsu.edu, anamika.dubey@wsu.edu.}
}

\maketitle

\begin{abstract}
    An electric power distribution system is operated in several distinct radial topologies by opening and closing of system’s sectionalizing and tie switches. The estimation of the system’s current operational topology is a precursor to implementing any optimal control actions (during normal operation) or restorative actions (during outage condition). This paper presents a mathematical programming approach to estimate the operational topology of a three-phase unbalanced power distribution system under both outage and normal operating conditions. Specifically, a minimum weighted least absolute value estimator is proposed that uses the line flow measurements, historical/forecasted load data, and ping measurements and solves a mixed-integer linear program (MILP) to estimate the operational topology and any outaged sections simultaneously. We thoroughly validate the accuracy of the proposed approach using IEEE 123-bus and a 1069-bus three-phase multi-feeder test system with the help of numerous simulations. It is observed that the approach performs well even at high percentages of measurement errors.		
\end{abstract}
	

\IEEEpeerreviewmaketitle
\vspace{-0.5 cm}
\section*{Nomenclature}
\vspace{-0.2 cm}
\addcontentsline{toc}{section}{Nomenclature}
\begin{IEEEdescription}[\IEEEusemathlabelsep\IEEEsetlabelwidth{$V_1,V_3$}]
    	\item[A. Sets]
		\item[$\mathcal{V}$] Set of buses in distribution system
		\item[$\mathcal{V}_c$] Set of buses connected to capacitors
		\item[$\mathcal{E}$] Set of distribution lines
		\item[$\mathcal{E}_s$] Set of distribution lines with switches 
		\item[$\mathcal{M}$] Set of measurements
		\vspace{0.1 cm}
		\item[B. Variables]
		\item[$\delta_{ij}$] Binary variables representing status of the switch between buses $i$ and $j$ 
		\item[$p_{Lj}^\phi$] Per-phase real load demand at bus $j$
		\item[$q_{Lj}^\phi$] Per-phase reactive load demand at bus $j$
		\item[$s_{Lj}^\phi$] Vector of per-phase real and reactive load demands at bus $j$, $[p_{Lj}^\phi; q_{Lj}^\phi]$
		\item[$P_{ij}^\phi$] Per-phase active power flow in line connecting buses $i$ and $j$
		\item[$Q_{ij}^\phi$] Per-phase reactive power flow in line connecting buses $i$ and $j$
		\item[$S_{ij}^\phi$] Vector of per-phase active and reactive power flows in line connecting buses $i$ and $j$, $[P_{ij}^\phi; Q_{ij}^\phi]$
		\item[$y_l$] Load section connectivity variable for section $l$
		\item[$y_{Cj}^{\phi}$] Per-phase capacitor connectivity variable for capacitor bank connected at bus $j$.
		\item[$S_{n_{p}}$] Sum of error in ping measurements
		\vspace{0.1 cm}
		\item[C. Parameters]
		\item[$\hat{p}_{Lj}^\phi$] Per-phase forecasted real load demand at bus $j$
		\item[$\hat{q}_{Lj}^\phi$] Per-phase forecasted reactive demand at bus $j$
		\item[$q_{Cj}^\phi$] Per-phase rated reactive power for capacitor at $j$
		\item[$\hat{s}_{Lj}^\phi$] Vector of per-phase forecasted values of real and reactive load demands at bus $j$; $[\hat{p}_{{Lj}}^\phi; \hat{q}_{{Lj}}^\phi]$
		\item[$\sigma_{{s_{j}^\phi}}^2$] Vector of variances in per-phase forecasted real and reactive load demands at bus $j$
		\item[$\hat{P_{ij}}^\phi$] Per-phase measured value of real power flow in line connecting buses $i$ and $j$ 
		\item[$\hat{Q_{ij}}^\phi$] Per-phase measured value of reactive power flow in line connecting buses $i$ and $j$
		\item[$\hat{S_{ij}}^\phi$] Vector of per-phase measured values of power flows in line connecting buses $i$ and $j$; $[\hat{P}_{ij}^\phi; \hat{Q}_{ij}^\phi]$
		\item[$\sigma_{{S_{ij}^\phi}}^2$] Vector of variances in per-phase real and reactive power flow measurements 
		\item[$\hat{y_j}$] Smart meter ping measurement of load at bus $j$
\end{IEEEdescription}

\vspace{-0.2 cm}

\section{Introduction}
\IEEEPARstart{{A}} typical distribution system includes tie-switches and sectionalizing switches and is operated in several distinct radial topologies. The problem of identifying the current operational topology from the given planned distribution system model using real-time measurements is termed as {\em{topology estimation problem}}. Since knowing the operational topology is the initial requirement for state-estimation (in conventional state estimation algorithms) and for implementing any optimal control action (during normal operation) or restorative actions (during outage condition), an incorrect estimate may lead to incorrect states and sub-optimal decisions.  Note that a vast body of literature is available on the real-time topology estimation algorithm for transmission systems using techniques such as weighted least squares state estimation, generalized state estimation, and least median of squares estimation \cite{monticelli1993modeling, da2016simultaneous, caro2010breaker, mili1996robust, phaniraj1991least, vempati2005topology, kekatos2012joint}. However, unlike transmission systems, a typical distribution system is radially operated and incurs unbalanced and lossy system conditions. Therefore, existing topology estimation algorithms from transmission systems literature are not directly applicable to distribution systems. In this paper, we present a scalable optimization-based approach to simultaneously estimate the operational topology and any outaged sections for an unbalanced three-phase power distribution system. 

Traditionally, a network topology processor (NTP) identifies the grid’s operational topology using the statuses of the switching devices by forming a linked list to capture the radial distribution structure \cite{korres2012state,he2001efficient}. The inputs to traditional NTP are the bus-switch model and the user-defined, measured, scheduled or normal status of the switching devices. The bus-switch model consists of all the branches of the system that can be opened/closed depending upon the corresponding switches being opened/closed. The traditional NTP processes the bus/switch model with the measured switch statuses and results in a bus-branch model of the system consisting of all system buses and `closed' branches of the network. Assuming full confidence in measured switch statuses, the traditional NTP simply forms a linked list that captures the radial distribution structure. It means that the traditional network topology processors (NTP) assume that the measured status of switches are accurate (i.e. they have no errors). Owing to frequent reconfigurations and manual changes by operators, the available switch status measurements may not reflect the actual switch statuses. Therefore, if the switch status measurements are incorrect, the radial topology identified by the ``traditional network topology processor'' will also be incorrect.

To include switch status errors, researchers have proposed generalized state estimation formulations for both transmission and distribution systems that simultaneously identify both continuous state variables and switch statuses \cite{monticelli1993modeling,da2016simultaneous,caro2010breaker, mili1996robust,alsac1998generalized, phaniraj1991least,vempati2005topology, kekatos2012joint,monticelli2000testing, lourenco2004bayesian, 1489505,singh2010recursive,compareGSE, korres2006substation,korres2012state}.
Unfortunately, due to nonlinear power flow models, the joint estimation  of the system’s  states  and  topology  leads  to  a  nonlinear  problem that  does  not  scale  well for large-scale unbalanced power distribution systems. In fact none of the above-mentioned methods have been demonstrated for a large unbalanced three-phase distribution system model; the following feeder models were used in the above referenced papers: IEEE 14-bus transmission test case, IEEE-24 bus transmission test system, RTS-96 bus transmission test system, 26-bus U.K. Generic Distribution System (UKGDS) model, and 17-bus artificial distribution system models. Furthermore, while in the transmission system, simplified/approximate power flow models can be used to simplify the generalized state estimation problem, the unbalances and losses in distribution systems prohibit one from simplifying power flow model for accurate state estimation. An unbalanced distribution system requires three-phase nonlinear power flow formulation that further increases the computational complexity making the generalized state estimation algorithms even more difficult to scale compared to the transmission systems. Therefore, for distribution systems, there is the need for a separate topology estimation stage where an approximate/linearized distribution power flow model can be used to identify the real-time operational tree/radial topology. Afterward, any standard state-estimation algorithm (with a non-linear power flow model) can be applied on the known topology to estimate system states.

To avoid non-linearity in topology estimation problems, in \cite{sharon2012topology}, authors establish a linear relation between real-time measurements and power injection statistics. Simulated real-time measurements for each topology are compared against the actual measurements to determine the switch statuses. Although the algorithm is linear in the number of topologies, evaluation of all possible operational topologies poses computational challenges. In \cite{sevlian2015distribution}, a polynomial complexity approximate Maximum a Posteriori (MAP) detector is proposed to identify the operational topology using only flow measurements by exhaustively searching for the correct operational topology. The worst-case scenario requires evaluation of all possible spanning trees and hence poses computational challenges for large systems. 
In \cite{zamani2015topology}, the distribution system topology is estimated using the residual of branch current distribution state estimation. The algorithm, however, also requires an exhaustive search of all possible topologies posing computational challenges for large systems. 

Another body of literature employs a data-driven approach. In \cite{cavraro2018power}, authors claim that a topology change leaves a signature in $\mu$PMU measurements and compare this signature with a library of signatures pre-determined for all possible topologies to detect distribution system topology. The algorithm assumes only one switch change at a time thus restricting the applicability for a real-world distribution system. Furthermore, for better performance on a larger system, the approach requires a large number of expensive $\mu$PMU measurement units. 
In related work, authors present data-driven methods to identify the distribution system connectivity model using several days of historical data \cite{liao2015distribution,deka2018topology}. These methods identify a {\em{static network connectivity model}} using a large set of voltage measurements. Alternatively, in this paper, we address a different problem where our objective is to estimate the {\em{time-varying}} operational topology of the distribution system with a known planning/connectivity model using real-time measurements.  

Unfortunately, the aforementioned literature on operational topology estimation assume that there are no outages and, therefore, cannot be extended to estimate the operational topology during outage conditions. For operational topology estimation during outages, our objective is to estimate both energized radial feeder and the outaged sections. The existing literature on outage detection assumes that the normal radial operational topology (without fault) is known \cite{jiang2016outage,sevlian2018outage, liao2016urban}. Starting with a radial topology, these methods identify outaged feeder sections and hence, are inapplicable when the current radial operational topology is not known.  

The existing literature on distribution grid topology estimation poses scalability concerns for large-scale unbalanced distribution systems, and cannot identify operational topology during outages when the normal radial operational topology is unknown. This calls for a generalized model for operational topology estimation that simultaneously takes the outage and normal operating conditions into account. In this paper, we propose a mixed-integer linear program (MILP) that simultaneously identifies the current operational topology for the unbalanced distribution grid and any outaged sections using line flow measurements, historical/forecasted load data, and ping measurements from a limited set of smart meters. The proposed framework can be thought of as an ``advanced network topology processor''.
The major contributions of this paper are as follows:
\begin{enumerate}[noitemsep,topsep=0pt,leftmargin=*]
	\item{\em Simultaneous Identification of Normal and Outaged Topology from Planning Model:} The proposed MILP formulation is a generalized framework that estimates the operational topology during both normal operation and outage condition without enumeration. The method incorporates the possibility of switching of system's legacy devices.		 
	\item{\em Incorporate Errors in both Continuous and Discrete Measurements:} The proposed formulation models both continuous and discrete measurements into one unified formulation. The measurement errors and their distributions are appropriately modeled and the estimation accuracy is calculated for varying percentages of measurement errors. 
    \item{\em Scalable MILP formulation for large-scale unbalanced distribution systems:} A computationally-tractable MILP formulation is proposed to solve the topology estimation problem for large-scale unbalanced distribution systems. To demonstrate the scalability, the approach is validated using a 1069-bus three-phase test system where on an average the operational topology is obtained within 30 sec.
\end{enumerate}

\begin{figure}[t]
	\centering
	\vspace{-0.2cm}
	\includegraphics[width=0.485\textwidth]{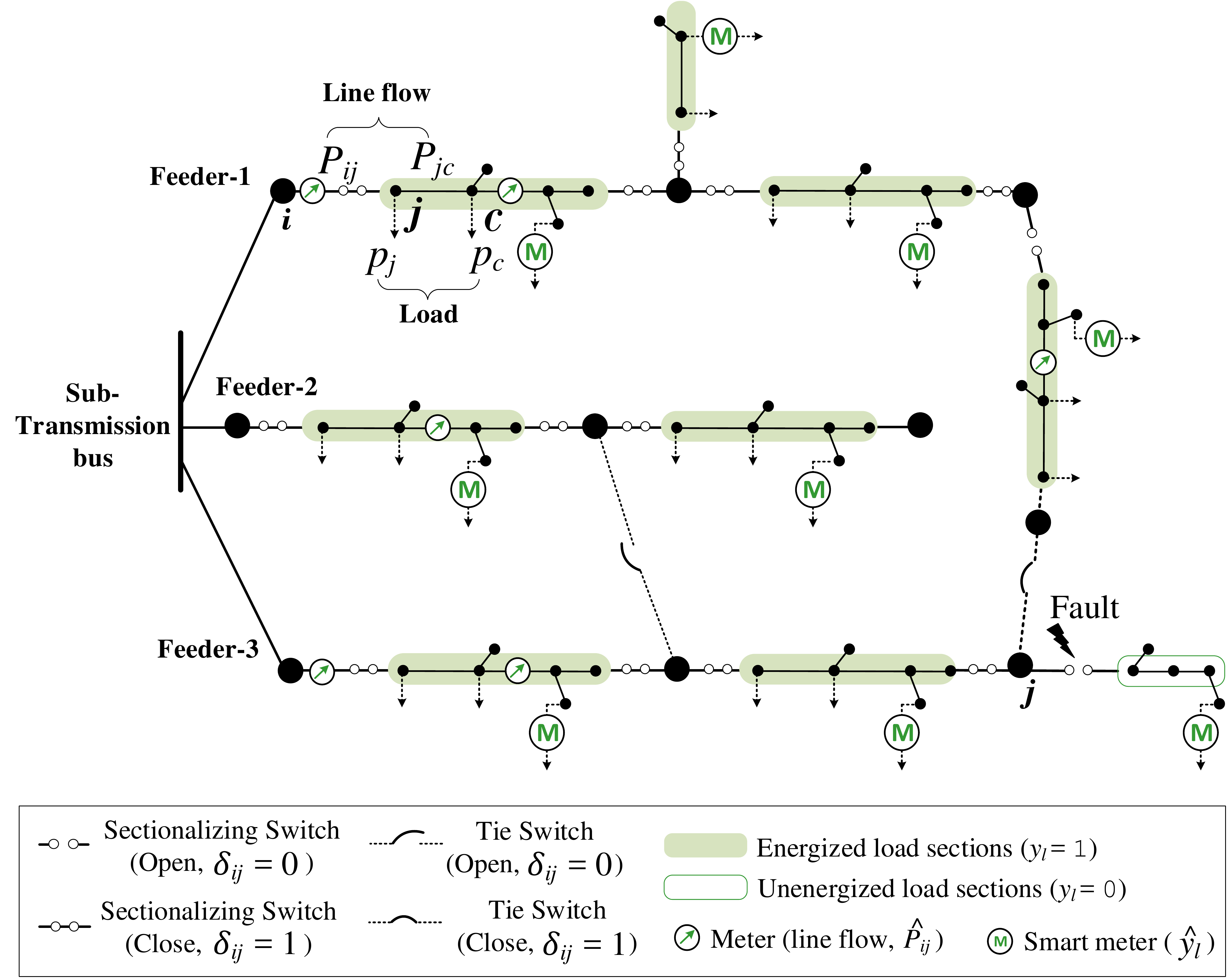}
	\vspace{-0.6cm}
	\caption{A simple distribution grid. The radial operational topology is dictated by the statuses of sectionalizing switches and tie switches. The operational topology changes as a result of reconfiguration during normal operation (spanning trees) or due to fault isolation during outages (subtrees).}
	\label{fig:1}
	\vspace{-0.9cm}
\end{figure}

\vspace{-0.3cm}	
\section{Measurement-based Topology Estimation}
The measurement-based topology estimation problem uses distribution system measurements to infer the operational topology of the distribution grid from the available planning model. First, we present a graph-theoretic framework to represent the three-phase power distribution system and associated topology estimation problem. Next, we detail the discrete and continuous variables to define the topology estimation problem and present the description of the measurement set. 

\vspace{-0.3cm}
\subsection{Graphical Representation of the Distribution System}
A distribution system is comprised of interconnected distribution feeders and can be operated in numerous radial operational topologies as shown in Fig. 1. Moreover, several normally open tie switches and normally closed sectionalizing switches are employed to facilitate the restoration efforts leading to numerous possible topologies during an outage. Let, the planning model be represented by graph $\mathcal{G}=(\mathcal{V},\mathcal{E})$ where $\mathcal{V}$ represents the set of buses and $\mathcal{E}$ is the set of edges representing all distribution lines and switches that can be open/closed. Let, the operational topology is represented by a tree $\mathcal{T} =(\mathcal{V_T},\mathcal{E_T}$), where, $\mathcal{E_T}\subset\mathcal{E}$ and $\mathcal{V_T}\subseteq\mathcal{V}$. Note that during normal operation, all buses are supplied and hence the operational topology, $\mathcal{T}$, is a spanning tree of the graph, $\mathcal{G}$. However, during an outage, the operational topology will be a subtree of the planned topology graph. This is because, in the event of a fault, multiple loads within a faulted load section will be isolated by the opening of respective switches to isolate the faulted load sections, resulting in a subtree.

\vspace{-0.4cm}

\subsection{Topology Estimation: Problem Description}
Theoretically, the topology estimation problem involves a combinatorial problem of searching for a feasible tree, $\mathcal{T}$, within the graph, $\mathcal{G}$, that satisfies the measurements. There are both continuous and discrete measurements available including line flow measurements (active and reactive branch power flows), historical or forecasted load measurements also called pseudo-measurements, and switch status and smart meter ping measurements. Note that if the switch statuses are known and assumed to be correct, the topology estimation is a trivial problem. Therefore, the traditional network topology processor assumes that switch measurements are unreliable and are not included in the measurement set.

A combinatorial search problem is not tractable for a large-scale system with a large number of possible tree configurations, especially, when outages are taken into consideration. This is because the enumeration-based methods require processing not only all spanning trees but also all subtrees of the original graph to detect operational topology during outages. We formulate a computationally tractable optimization model to identify switch statuses by minimizing the weighted least absolute error in measured and predicted values of system variables subject to power flow constraints. Mathematical expressions for predicted line flows are obtained using a three-phase linearized power flow model coupled with binary switch variables. This results in an MILP model that estimates the switch statuses while incorporating the measurement errors.

\vspace{-0.4cm} 
\subsection{System Variables}
As stated in the problem description, switch statuses are the unknown variables of interest. To aid us in finding the switch statuses, we define load demand and power flow variables associated with each load and line in the distribution system. Also, to include possible outages, we define binary variables indicating outaged load sections. 
	
\subsubsection{Switch Status Variable} A binary variable $\delta_{ij}=\{0,1\}$ is associated with each line section including switches, where $\delta_{ij}=1$ implies that switch connecting buses $i$ and $j$ is closed, while $\delta_{ij}=0$ implies that the corresponding switch is open. The switch status variables are used to formulate power flow equations and connectivity model for the distribution system. 
	
\subsubsection{Load Section Connectivity Variable}
A load section is comprised of a minimum set of loads that must be disconnected to isolate a fault. During an outage, one/more load sections are disconnected from the main feeder by automated operation of protection devices. All loads within a load section are disconnected during a fault within that section.  
We define a binary variable, $y_{l}$ corresponding to each load section, where, $y_{l} = 0$ implies load section $l$ is outaged and vice versa. The estimation of load section connectivity variables leads to the identification of both operational topology and outaged sections for the given distribution system planning model. 
    
\subsubsection{Power Flow and Load Demand Variables} Per-phase active and reactive load demands at bus $j$ are represented by $p_{Lj}^\phi$ and $q_{Lj}^\phi$. Similarly, per-phase active and reactive power flows in line connecting buses $i$ and $j$, directed from $i$ to $j$, are represented by $P_{ij}^\phi$ and $Q_{ij}^\phi$, respectively.
    
\subsubsection{Capacitor Status Variables} Single-phase and three-phase capacitor banks are usually connected to the distribution system. Let, $y_{Cj}^\phi$ be the binary variable indicating status (ON/OFF) of the capacitor bank at bus $j$ in phase $\phi$, where, $y_{Cj}^\phi=1$ implies the capacitor bank is ON and vice-versa.
    
\vspace{-0.4cm} 

\subsection{System Measurements}
The measurements required for the proposed formulation include: (1) forecasted active and reactive load demands of all loads in the system, (2) active and reactive power flow measurements at a few selected distribution lines, (at least one flow measurement from each fundamental cycle \cite{sevlian2015distribution}), and (3) smart meter ping measurements from at least one load in each load section. We adopt the conventional notation of real and pseudo-measurements \cite{schweppe1974static} to categorize the telemetered flow measurements and forecasted load measurements, respectively. Note that the forecasted load measurements can be replaced by corresponding AMI data, if available. Measurement devices and forecasting techniques have errors that cannot be avoided. Therefore, measurements are modeled as random variables. The following section details the error models for different measurement variables and corresponding distributions for the associated measurement random variables. 

\subsubsection{Power Flow Measurements} Active and reactive power flow measurements for phase $\phi$ of the line between buses $i$ and $j$, directed from $i$ to $j$, are represented as random variables  $\hat{P}_{ij}^\phi$ and $\hat{Q}_{ij}^\phi$, respectively. The corresponding errors in these measurements are modeled as Gaussian distribution with zero mean and prespecified variance, i.e. $e(\hat{P}_{ij}^\phi) \sim \mathcal{N}(0,\sigma^2_{{P}_{ij}^\phi})$ and  $e(\hat{Q}_{ij}^\phi)=\mathcal{N}(0,\sigma^2_{{Q}_{ij}^\phi})$, respectively. Since these are telemetered measurements, they have low errors in the range of 1\% to 5\%. 

The power flow measurements are related to the power flow variables. Notice that, $\hat{P}_{ij}^\phi$ = ${P}_{ij} + e(\hat{P}_{ij}^\phi)$ and $\hat{Q}_{ij}^\phi$ = ${Q}_{ij} + e(\hat{Q}_{ij}^\phi)$. Therefore, the random variables corresponding to flow measurements follow the following distributions: $\hat{P}_{ij}^\phi \sim \mathcal{N}(P_{ij}^\phi,\sigma^2_{{P}_{ij}^\phi})$ and $\hat{Q}_{ij}^\phi \sim \mathcal{N}(Q_{ij}^\phi,\sigma^2_{{Q}_{ij}^\phi})$.

\subsubsection{Forecasted (Pseudo) Load Demand Measurements} Active and reactive load demand measurements at bus $j$ in phase $\phi$ are represented by random variables $\hat{p}_{Lj}^\phi$ and $\hat{q}_{Lj}^\phi$ respectively. The corresponding errors in the load measurements are modeled as Gaussian random variables with zero mean and prespecified variance i.e.  $e(\hat{p}_{Lj}^\phi) \sim \mathcal{N}(0,\sigma^2_{{p}_{Lj}^\phi})$ and $e(\hat{q}_{Lj}^\phi) \sim \mathcal{N}(0,\sigma^2_{{q}_{Lj}^\phi})$. As the load values are forecasted and used as pseudo measurements, they may incur high errors. 

The load demand measurements are related to the load demand variables. Notice that, $\hat{p}_{Lj}^\phi$ = ${p}_{Lj} + e(\hat{p}_{Lj}^\phi)$ and $\hat{q}_{Lj}^\phi$ = ${q}_{Lj} + e(\hat{q}_{Lj}^\phi)$. Therefore, the random variables corresponding to load demand measurements follow the following distributions: $\hat{p}_{Lj}^\phi \sim \mathcal{N}(p_{Lj}^\phi,\sigma^2_{{p}_{Lj}^\phi})$ and $\hat{q}_{Lj}^\phi \sim \mathcal{N}(q_{Lj}^\phi,\sigma^2_{{q}_{Lj}^\phi})$.

\subsubsection{Smart Meter Ping Measurements} 
During an outage, section(s) of the distribution grid is(are) isolated and hence the operational topology will be a subtree of $\mathcal{G}$ instead of a spanning tree. Therefore, we require additional measurements carrying the information regarding outaged sections. Existing outage management systems (OMS) include smart meter ping measurements to detect the connectivity status of a load. A smart meter, when pinged, sends out its load consumption, if energized, and does not respond when it is disconnected due to an outage. We consider smart meter communication as an additional measurement for the topology estimation problem. The smart meter ping measurement is given by $\hat{y}_j$ where, if the smart meter connected at $j^{th}$ load bus responds to the ping request, $\hat{y}_j=1$, implying that the load is energized; otherwise, $\hat{y}_j=0$, implying load is disconnected. 

We consider error in ping measurements, where, error is modeled using Bernoulli distribution i.e. $e(\hat{y}_j) \sim Bernoulli(q)$, where $q$ is the probability that the ping measurement is inaccurate, i.e. with $q$ probability, the difference between the ping measurement and the actual load section energization variable is 1 and with $(1-q)$ probability the difference is 0. This implies that $(\hat{y}_j - {y}_l) \sim Bernoulli(q)$, where load connected at bus $j$ is supplied by the load section $l \in L$. Let, a total of $l_{p}$ number of loads are pinged for each load section $l \in L$. Then total number of ping measurements are given by $n_{p} = \sum_{l = 1}^L l_{p}$. 

Next, we identify the distribution of the sum of errors in smart meter ping measurements. Note, that the sum of $n$ identically distributed Bernoulli random variables with parameter $k$ is given by a binomial distribution with parameters $n$ and $k$ i.e. $B(n,k)$. Therefore, the following statement (1) is true for the sum of error in ping measurements, $S_{n_{p}}$. Also, the mean and variance of $S_{n_{p}}$ is $n_{p}q$ and $n_{p}q(1-q)$.    
\begin{equation}
    S_{n_{p}} = \sum_{l = 1}^L \left(\sum_{j =1}^{l_{p}}|\hat{y}_j - {y}_l|\right) \sim B(n_{p},q)
\end{equation}

\vspace{0.3cm}
\vspace{-0.7cm}

\subsection{Topological Observability}
A brief discussion on conditions to ensure topological observability is presented here. Observability, as per control theory, is a measure of how well the internal states of a system can be inferred from the knowledge of the outputs. In similar terms, topological observability is defined as a measure of how well the system topology can be inferred (identified) from the knowledge of measured flows. The minimum number of measurements that ensure topological identifiability, during normal operation (without outages), with error-free measurements is stated in \cite{sevlian2015distribution}. The condition claims that the measurement set should be such that each possible spanning tree of the graph differs at least by one flow measurement to make the respective tree identifiable. This leads to the requirement of one flow measurement along each fundamental cycle \cite{sevlian2015distribution}. 

Following a similar line of thoughts, during an outage, the measurement set should be able to distinguish not only all spanning trees but also the sub-trees of the graph. Recall that the isolation of load sections results in different outaged topologies/sub-trees. Smart meter probing provides an obvious and economical approach to detect isolated load sections of the distribution system. To ensure each sub-tree is identifiable, at least one smart meter ping is required from each load section in addition to the flow measurements along the fundamental cycles. But due to the error in smart meter ping measurements and the associated cost of pinging a smart meter {\cite{tram2008technical}}, in this paper, we assume 10\% smart meter ping measurements from each load section (that can be erroneous) and one flow measurement per fundamental cycle are available. 

\textit{Practical Considerations regarding Smart Meter Ping Measurements for Topology Estimation:} 
The capability to ping smart meters using the existing AMI has been used in the existing literature on outage management systems (OMS) for fault location and isolation functions; see references \cite{jamali2017fault, majidi2014fault,itron}. In fact, a modern OMS includes several AMI functions including outage notification (via smart meter last gasp measurements), outage confirmation (via smart meter pings), system restoration (AMI reports restoration), and restoration verification (by checking the status using smart meter ping). Therefore, under the current conditions, it is realistic to ping 10\% smart meters for topology estimation. 

\begin{itemize}[leftmargin=*]
	\item{\em Asynchronous Measurements:} One critical aspect when including  smart meter pings in topology estimation problem is their time alignment with other measurements and pseudo-measurements. Note that asynchronous smart meter ping measurements may theoretically affect the accuracy of the proposed topology estimation algorithm. However, practically that is not the case. Since smart meter pings are operator initiated, operators have the autonomy to ping the meters at the convenient time steps to intentionally synchronize the smart meter pings with the other continuous measurements. Furthermore, the proposed approach performs well even with the errors in ping measurements; these errors could represent a few asynchronous smart meter pings. 
	\item{\em Bernoulli Distribution for Ping Measurements:} A Bernoulli distribution is a discrete probability distribution of a random variable which takes the value $1$ with probability $q$ and the value $0$ with probability $(1-q)$. Basically, it models any single experiment with a yes-no question that results in the Boolean-valued outcome ($0/1$). When modeling error in ping measurement i.e., $e(\hat{y}_j) = |(\hat{y}_j - {y}_l)|$ as Bernoulli distribution, we are asking the question whether the measured and true status of the smart meters are the same. The outcome is Boolean-valued ($0/1$) that depends upon the error in measuring or reporting smart meter pings.
\end{itemize}

\vspace{0.1cm}

\textit{Discussion on problems related to bad data:} As any LAV estimator, the proposed algorithm is susceptible to bad data errors. If enough redundant real-time measurements are available, related methods from transmission systems literature can be employed for bad data detection and elimination prior to implementing topology estimation algorithm \cite{singh2005topology}. However, recognizing that the distribution systems have relatively fewer measurements compared to the transmission networks, the topology estimation algorithm needs to be designed keeping in mind the scarcity of measurements in the distribution systems. If there are fewer measurements, a bad data detection and elimination problem is extremely difficult to solve. This is because eliminating even a single measurement (due to bad data error) can render the topology unobservable (see discussion regarding topological observability in Section II.E). Specialized algorithms are required for bad data detection and elimination in distribution systems with measurement scarcity. This, although an interesting direction for future research, is outside the scope of this work.
	
\vspace{-0.3cm}

\section{Problem formulation}
This section describes the problem formulation for distribution system topology estimation problem. First, we describe a maximum likelihood estimator (MLE) for topology estimation. Next, the MLE problem is reformulated into a computationally tractable optimization model. 

\vspace{-0.3cm} 

\subsection{Description of Likelihood Function}
Let, there be $|N|$ possible topologies in $\mathcal{G}$ and $\Delta_k$ be the set of all switch status variables corresponding to $k^{th}$ topology. Also let, the measurement set, $\mathcal{M}$ = $\{\mathcal{\hat{P}},\mathcal{\hat{Q}}, \hat{p},\hat{q},\hat{y}\}$, where,  $\mathcal{\hat{P}}$ and $\mathcal{\hat{Q}}$, be the set of all active and reactive power flow measurements; $\hat{p}$ and $\hat{q}$, be the set of all active and reactive load demand measurements; and  $\hat{y}$ be the set of all smart meter ping measurements. Then, the likelihood of observing a given topology, $\Delta_k$, based on the measurement set, $\mathcal{M}$, is defined as $\mathcal{L}(\Delta_k|\mathcal{M})$, for all $k\in\{1 ... |N|\}$. The aim is to estimate the most likely topology, $\Delta_k$, using the erroneous measurement set $\mathcal{M}$ by maximizing the likelihood function $\mathcal{L}(\Delta_k|\mathcal{M})$.

The optimization model requires an analytical expression for $\mathcal{L}(\Delta_k|\mathcal{M})$. It is difficult to analytically characterize the probabilistic model for categorical variables, i.e. $\Delta_k$. Note that each topology will induce a different set of system variables, $\mathcal{X}_{\Delta_k}$. Therefore, a new likelihood function, $\mathcal{L}(X_{\Delta_k}|\mathcal{M})$, can be defined that measures the likely system variables from the given measurement set and can be obtained using the error models for different measurement variables in $X_{\Delta_k}$. The problem objective is to obtain the most likely system variables by maximizing the expression for $\mathcal{L}(X_{\Delta_k}|\mathcal{M})$. The resulting topology induced by the most likely system's variables is the most likely operational topology.

The variables set, $X_{\Delta_k}$, includes both continuous ($\mathcal{\hat{P}},\mathcal{\hat{Q}},\hat{p},\hat{q}$) and discrete variables ($\hat{y}$), thus making the joint characterization of the likelihood function difficult. One approach is to maximize the likelihood of both continuous and discrete variables is by defining a multi-objective optimization problem that maximizes a weighted sum of the respective likelihood functions. Unfortunately, tuning the weight parameters for the continuous and categorical variables is difficult to generalize. In the proposed approach, we formulate a single objective optimization problem where the estimation problem maximizes the likelihood of observing only continuous random variables. The probabilistic representation of discrete random variables is modeled in constraint formulation. The objective is to estimate most likely power flow and load demand variables while constraining the probability of observing discrete variables to lie within a pre-specified interval based on the known error distribution.

\vspace{-0.3cm}

\subsection{Objective Function Formulation}
The MLE problem for Gaussian random variables is equivalent to the {\em{method of least squares}}. Therefore, the objective of maximizing the likelihood function for continuous Gaussian random variables leads to the problem of minimizing the squared errors between system measurements and respective variables weighted according to measurement variances as defined in (\ref{eq:nlnobj}). 
The topology that has loads and flows closest to the measured values is the correct operational topology.   
\begin{equation}\label{eq:nlnobj}
		\small
		\text{Min}. \displaystyle \sum_{\phi \in \{a,b,c\}}\Bigg({\displaystyle \sum_{j \in \mathcal{I}} \bigg(\frac{(\hat{s}_{Lj}^\phi-s_{Lj}^\phi)}{(\sigma_{s_{Lj}^\phi})}\bigg)^2+\displaystyle \sum_{ij \in \mathcal{B}} \bigg(\frac{(\hat{S}_{ij}^\phi-S_{ij}^\phi)}{(\sigma_{S_{ij}^\phi})}\bigg)^2\Bigg)}
\end{equation}
where, $\mathcal{I}$ is the set of pseudo load measurements and $\mathcal{B}$ is the set of line flow measurements. Note that load and flow measurements are per-phase complex quantities; $\hat{s}_{Lj}^\phi$, $s_{Lj}^\phi$, $\hat{S}_{ij}^\phi$, $S_{ij}^\phi$ are vectors of corresponding active and reactive power components (see Nomenclature).
Note that the least-squares problem leads to a nonlinear objective function. When coupled with binary variables (in constraints), this results in an MINLP problem that is difficult to scale. Here, we take a cue from Least Absolute Value state estimator that has been proved to be accurate for state estimation in past \cite{singh1994weighted}. Following this, we define a linearized objective function in (\ref{eq:obj}) that minimizes the weighted absolute errors instead of squared errors.   
\begin{equation}\label{eq:obj}
	\small
	\text{Min}. \displaystyle \sum_{\phi \in \{a,b,c\}}\Bigg({\displaystyle \sum_{j \in \mathcal{I}} \left|\frac{(\hat{s}_{Lj}^\phi-s_{Lj}^\phi)}{(\sigma_{s_{Lj}^\phi})}\right|+\displaystyle \sum_{ij \in \mathcal{B}} \left|\frac{(\hat{S}_{ij}^\phi-S_{ij}^\phi)}{(\sigma_{S_{ij}^\phi})}\right|\Bigg)}
\end{equation}
 
The absolute value in the objective function is still nonlinear. However, this can be easily linearized by introducing new variables (vectors) $a$ and $b$ such that
$\mid s_{Lj}^\phi-\hat{s}_{Lj}^\phi\mid=a(j)$ and $\mid S_{ij}^\phi-\hat{S}_{ij}^\phi)\mid=b(i)$ and by introducing the following additional inequality constraints in the problem formulation:

\vspace{-0.4cm}
    
\begin{small}
\begin{eqnarray}\label{eq:abs}
 s_{Lj}^\phi-\hat{s}_{Lj}^\phi\leq a(j), \hspace{0.3cm} &\text{and}& \hspace{0.3cm}
-(s_{Lj}^\phi-\hat{s}_{Lj}^\phi)\leq a(j)\\
S_{ij}^\phi-\hat{S}_{ij}^\phi \leq b(i), \hspace{0.3cm} &\text{and}& \hspace{0.3cm}
-(S_{ij}^\phi-\hat{S}_{ij}^\phi) \leq b(i)
\end{eqnarray}
\end{small}
\vspace{-1cm}	
\subsection{Constraint Formulation}
We identify three types of constraints: 1) power flow, 2) connectivity, and 3) error bounds on ping measurements.  
\subsubsection{Power flow Constraints}
A three-phase linearized AC power flow model for the unbalanced distribution system, proposed in \cite{low2014convex}, is used. The linearized three-phase formulation is sufficiently accurate for topology estimation \cite{sevlian2015distribution}. Note that a linearized model ignores power losses. Therefore, the proposed approach can not distinguish between two topologies that only differ by losses in the flow measurements; however, this rarely happens for a practical distribution system \cite{sevlian2015distribution}. 
\begin{eqnarray}
v_j  = v_i - (S_{ij}z_{ij}^H+z_{ij}S_{ij}^H) + z_{ij}l_{ij} z_{ij}^H \\
\text{diag}(S_{ij} - z_{ij}l_{ij}) =  \sum_{k:j \rightarrow k}{\text{diag}(S_{jk})}  + s_{L,j}  \\
\left[
  \begin{array}{cc}
    v_i & S_{ij} \\
    S_{ij}^H & l_{ij} \\
  \end{array}
\right] = \left[
  \begin{array}{c}
    V_i\\
    I_{ij}\\
  \end{array}
\right]
\left[
  \begin{array}{cc}
    V_i\\
    I_{ij}\\
  \end{array}
\right]^H \\
\left[
  \begin{array}{cc}
    v_i & S_{ij} \\
    S_{ij}^H & l_{ij} \\
  \end{array}
\right] : - \text{Rank-1 PSD Matrix}
\end{eqnarray}

First, a three-phase power flow formulation for a radial system based on branch flow relationship given in \cite{gan2014convex} is detailed. Let, there be directed graph $\mathcal{G} = (\mathcal{N}, \mathcal{E})$ where $\mathcal{N}$ denotes set of buses and $\mathcal{E}$ denotes set of lines. Each line connects ordered pair of buses $(i,j)$ between two adjacent nodes $i$ and $j$. Let, $\{a,b,c\}$ denotes the three phases of the system; $V_i := [V_i^{\phi}]_{\phi \in \{a,b,c\}}$ be the three phase voltage at node $i$; $I_{ij}^{\phi}$ be the $\phi$ phase current for line $(i,j)$ and define $I_{ij} := [I_{ij}^{\phi}]_{\phi \in \{a,b,c\}}$; and $z_{ij}$ be the phase impedance matrix. Then (6)-(8) represent nonlinear three-phase power flow equations.  

Next, we describe the linear approximation for the three-phase branch flow equations obtained after approximating the nonlinear power flow model described in (6)-(9). The linear approximation assumes the branch power loss is relatively smaller as compared to the branch power flow \cite{gan2014convex}. Specifically, the impact of power loss on active and reactive power branch flow equations and on voltage drop equations is ignored. The linearized AC branch flow equations are shown in (10)-(11). Here, (10) corresponds to linearized active power flow, and (11) represents linearized reactive power flow equations. Note that we do not include voltage equations in topology estimation problem as they provide little inference in identifying the operational topology. This is because, in distribution systems, tight control of  grid voltages results in approximately similar nodal voltage magnitudes for different feeder topologies \cite{sevlian2015distribution}.
\begin{equation}\label{eq:p1}
\small
       \sum_{(i\rightarrow j) \in \mathpzc{E}} \textit{P}_{ij}^{\phi} = \ \textit{p}_{Lj}^{\phi} + \sum\limits_{\substack{(j\rightarrow c) \in  \mathpzc{E}, i\neq c}}\textit{P}_{jc}^{\phi} \hspace{0.2 cm} \forall j\in \mathcal{V}
\end{equation}
\begin{equation}\label{eq:q1}
\small
       \sum_{(i\rightarrow j) \in \mathpzc{E}} \textit{Q}_{ij}^{\phi} = \ \textit{q}_{Lj}^{\phi} + \sum\limits_{\substack{(j\rightarrow c) \in  \mathpzc{E}, \\i\neq c}} \textit{Q}_{jc}^{\phi} \hspace{0.2 cm} \forall j\in \mathcal{V} \cap \mathcal{V}_c 
\end{equation}

The linearized power flow equations are only valid if the line and the corresponding loads are energized. To formulate topology-constrained power flow model, the branch flow equations are coupled with switch status, load section connectivity and capacitor connectivity variables. Note that $\delta_{ij}= 1$, if $(i\rightarrow j) \in \mathcal{E}$\textbackslash $\mathcal{E}_s$. Equations (\ref{eq:p})-(\ref{eq:c}) represent three-phase unbalanced linearized power flow equations coupled with switch status variable $\delta_{ij}$, load section connectivity variables $y_l$, where load $j$ belongs to load section $l \in L$ and capacitor connectivity variables $y_{Cj}^{\phi}$. Here (\ref{eq:p})-(\ref{eq:c}) define power flow constraints that must be satisfied for each energized line.

\vspace{-0.4cm}

\begin{equation}\label{eq:p}
\small
       \sum_{(i\rightarrow j) \in \mathpzc{E}}\delta_{ij} \textit{P}_{ij}^{\phi} = y_l \ \textit{p}_{Lj}^{\phi} + \sum\limits_{\substack{(j\rightarrow c) \in  \mathpzc{E}, i\neq c}} \delta_{jc} \textit{P}_{jc}^{\phi} \hspace{0.2 cm} \forall j\in \mathcal{V}
\end{equation}
\begin{equation}\label{eq:q}
\small
       \sum_{(i\rightarrow j) \in \mathpzc{E}}\delta_{ij} \textit{Q}_{ij}^{\phi} = y_l \ \textit{q}_{Lj}^{\phi} + \sum\limits_{\substack{(j\rightarrow c) \in  \mathpzc{E}, \\i\neq c}}     \delta_{jc} \textit{Q}_{jc}^{\phi} \hspace{0.2 cm} \forall j\in \mathcal{V} \cap \mathcal{V}_c 
\end{equation}
\vspace{-0.4cm}

For buses with capacitor banks, i.e. $j \in \mathcal{V}_c$, the reactive power flow equation need to be coupled with capacitor bank's switching status as defined in (\ref{eq:c}).
\begin{equation}\label{eq:c}
\small
        \sum_{(i\rightarrow j) \in \mathpzc{E}}\delta_{ij} \textit{Q}_{ij}^{\phi} = y_l \ (\textit{q}_{Lj}^{\phi} - y_{Cj}^{\phi} \ \textit{q}_{Cj}^{\phi}) + \sum\limits_{\substack{(j\rightarrow c) \in  \mathpzc{E},\\ i\neq c}} \delta_{jc} \textit{Q}_{jc}^{\phi} \hspace{0.0 cm} \forall j\in \mathcal{V}_c
\end{equation}

Note that constraints (\ref{eq:p})-(\ref{eq:c}) involve product of variables and are linearized using big-M method. 

\subsubsection{Connectivity Constraints}
This section defines different connectivity constraints that must be satisfied to obtain a feasible radial operational topology. 

\begin{itemize}[leftmargin=*]
	\item{\em Cycle Constraints:}
	The cycle constraints are to ensure that the grid operates in a radial topology with no cycles. Let, $\mathcal{C}$ be the set of all possible cycles in the grid. Then, (\ref{eq:cyc}) is imposed on each cycle $k \in \mathcal{C}$ to ensure that all the loads in the network are supplied radially without forming loops.	
	\begin{equation}\label{eq:cyc}
	\small
		\sum_{ij\in \mathcal{C}(k)}\delta_{ij}\leq n_{sw}(k)-1; \hspace{0.5 cm} k=1....m
	\end{equation}		
    where, $n_{sw}(k)$ is the number of switches in cycle $k$.
	\item{\em  Load-switch Connectivity Constraints:}
	These constraints ensure a feasible outage topology by establishing relationships between power flows in a load section and the statuses of all corresponding switches during outages.  
	\begin{itemize}
		\item A load $s_{Lj}^\phi$ at bus $j$ in phase $\phi$ downstream from a single switch, $\delta_{ij}$, is connected to the power supply if and only if the corresponding switch is closed as described in (\ref{eq:bin}). Note that for both single-phase and three-phase loads belonging to load section $l \in L$, we specify a single binary variable $y_l$ to represent the load connectivity status.
		\vspace{-0.2cm}
		\begin{equation}\label{eq:bin}
		\small
			y_l=\delta_{ij} 
		\end{equation}
		\item A load $s_{Lj}^\phi$ at bus $j$ in phase $\phi$ that can be supplied by a total of $n$ switches, $\{\delta_{ij}$....$\delta_{nj}\}$, due to open-loop distribution system configuration, is connected to the power supply if any one of the switches is closed. This constraint is defined using the set of equations in (\ref{eq:loop}). 
		\vspace{-0.1cm}
		\begin{equation}\label{eq:loop}
		\small
			 y_l \leq \sum_{i = 1}^n\delta_{ij} \hspace{0.2cm}  \text{ and } \hspace{0.2cm} y_l \geq \delta_{ij} \hspace{0.2cm} \forall{i} \in \{1...n\}
		\end{equation}
	\end{itemize}
	\item{\em  Ping Variable Constraints:}
	If a smart meter responds to a ping request, the corresponding load cannot be in outage i.e., if $\hat{y}_j=1$, then $y_l=1$ where, load at bus $j$ is connected to load section $l\in L$ as defined in (\ref{eq:smart}).
	\begin{equation}\label{eq:smart}
	\small
	y_l \geq \hat{y}_j
	\end{equation}
	\item{\em Switch-flow Constraints:}
    If a switch $\delta_{ij}$ is open, then power flow, $\textit{S}_{ij}^\phi$, through the switch must be equal to zero. Otherwise the flow will be unconstrained as defined in (\ref{eq:swf}).  
    \begin{equation}\label{eq:swf}
    	\small
    		-M\delta_{ij} \leq \textit{S}_{ij}^\phi \leq M\delta_{ij}; \hspace{0.5 cm} M >> 0
    	\end{equation}
    	where, $M$ is a large positive number.
\end{itemize}
	
\begin{figure}[t]
	\centering
	\includegraphics[width=0.46\textwidth]{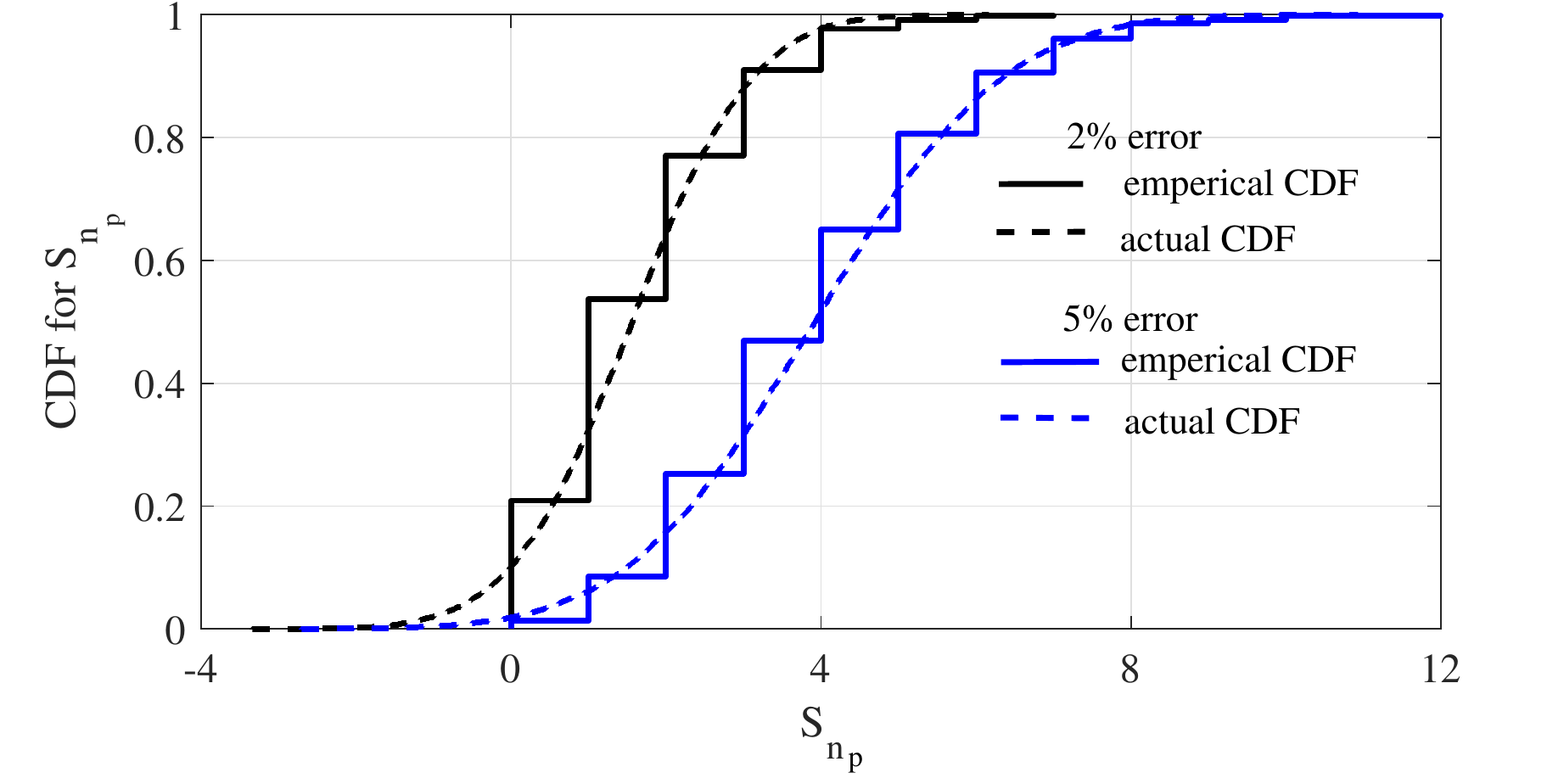} 
	\vspace{-0.4cm}
	\caption{Comparisons of CDF sum of Bernoulli variables and its Gaussian approximation for 2\% and 5\% error in ping measurements.}
	\label{fig:2}
	\vspace{-0.9cm}
\end{figure}
			
\subsubsection{Error Bounds in Smart Meter Ping Measurements}
The objective function defined in (3) does not include the error in discrete smart meter ping measurements. As described before, unlike continuous variables, we model the errors in discrete measurements as constraints. The idea is to ensure that the estimated load section variables appropriately characterize the possible errors in smart meter pings. 
	
As detailed in Section II.D, for a sufficiently large number of ping measurements, the sum of absolute errors in ping measurements, $ S_{n_{p}} $, can be modeled as Binomial distribution, see (1), with mean $ \mu_e = n_{p}q$ and variance $\sigma_e = \sqrt{n_{p}q(1-q)}$. Note that $S_{n_{p}}$ can be approximation as a Gaussian distribution even for modest sample size. We demonstrate this observation using Fig. \ref{fig:2} where the PDF for $S_{n_{p}}$ for the multi-feeder test system having a total of $44$ load sections 
is obtained for 2\% ($q = 0.02$) and 5\% ($q = 0.05$) errors in ping measurements. The sample distribution is compared with a Gaussian distribution obtained using same mean and variance parameters ($\mu_e = n_{p}q$ and $\sigma_e = \sqrt{n_{p}q(1-q)}$). 
	
Next, we derive constraints that ensure a 99.99\% confidence in estimating $S_{n_{p}}$ for a given measurement set. For Gaussian random variables, a 99.99\% confidence-level corresponds to 5$\sigma$ spread around the sample mean. Therefore, the bounds on $S_{n_p}$ are defined in (\ref{eq:ping}) which imply that the sum of ping errors are bounded by $5\sigma$ spread of the associated distribution. 
	\begin{equation} \label{eq:ping}
	\small
	min(0,\mu_e-5\sigma_e) \leq  S_{n_{p}} \leq \mu_e+5\sigma_e
	\end{equation}
	
The resulting optimization problem is an MILP with the objective function defined in (\ref{eq:obj}) subject to (\ref{eq:abs}), (5), (\ref{eq:p})-(\ref{eq:ping}).  

Note that here we assume that a smart meter ping can respond inaccurately during both conditions when connected as well as when outaged. However, when smart meter is outaged it will not respond and $\hat{y}_j$ will be always 0, therefore, implying that ping errors have different distribution when $y_l$ is 0 or 1. We argue that a Bernoulli distribution is a valid assumption even with the case that outraged smart meter cannot respond erroneously for the following two reasons: 1) in a realistic outage scenario, there will be a smaller fraction of smart meters that will be disconnected and hence the distribution for $S_{n_{p}}$ will be dominated by the error in pings from connected smart meters; 2) even when actual sum of ping errors is lower as disconnected smart meters pings are never erroneous, the constraint formulation using $S_{n_{p}}$ for topology estimation given in (20) does not remove the correct topology from the search space of the proposed MILP algorithm. This can be observed by calculating the new constraint for the sum of error in ping measurement due to only those smart meters that are connected and comparing that to (20).	

\vspace{0.1cm}
\textit{Discussion on Accuracy of Secondary Feeder Models:} The proposed framework requires an accurate planning model including secondary feeder models particularly phase association for secondary transformer and customer loads for the underlying distribution system. For a distribution system, the primary feeder model (including connectivity and equipment models) are reasonably accurate. However, the secondary feeder models are known to be inaccurate. The secondary model inaccuracies include phase assignment errors for secondary loads, incorrect transformer phase connections, and incorrect triplex/secondary line codes, etc. Driven by the need to leverage secondary feeder data, lately, the secondary feeder model estimation problem has received significant attention. Several data-driven and model-based methods have been proposed to identify the accurate phase assignment for the customer loads, accurate transformer connections, and secondary line parameters \cite{arya2011phase, short2012advanced, 8586483, park2018exact}.  Any of these methods can be used to estimate or improve secondary feeder models for the underlying distribution system. Afterward, the proposed approach can be used to model the topology estimation problem. We would like to emphasize that solving model estimation problem is outside of the scope of this work. Usually, model and parameter estimation algorithms are implemented at an earlier stage to obtain an accurate planning model for the underlying system. Topology estimation and state estimation problems are operational problems that are implemented on a known and reasonably accurate system model.
		
\begin{figure}[t]
	\centering
	\includegraphics[width=0.45\textwidth]{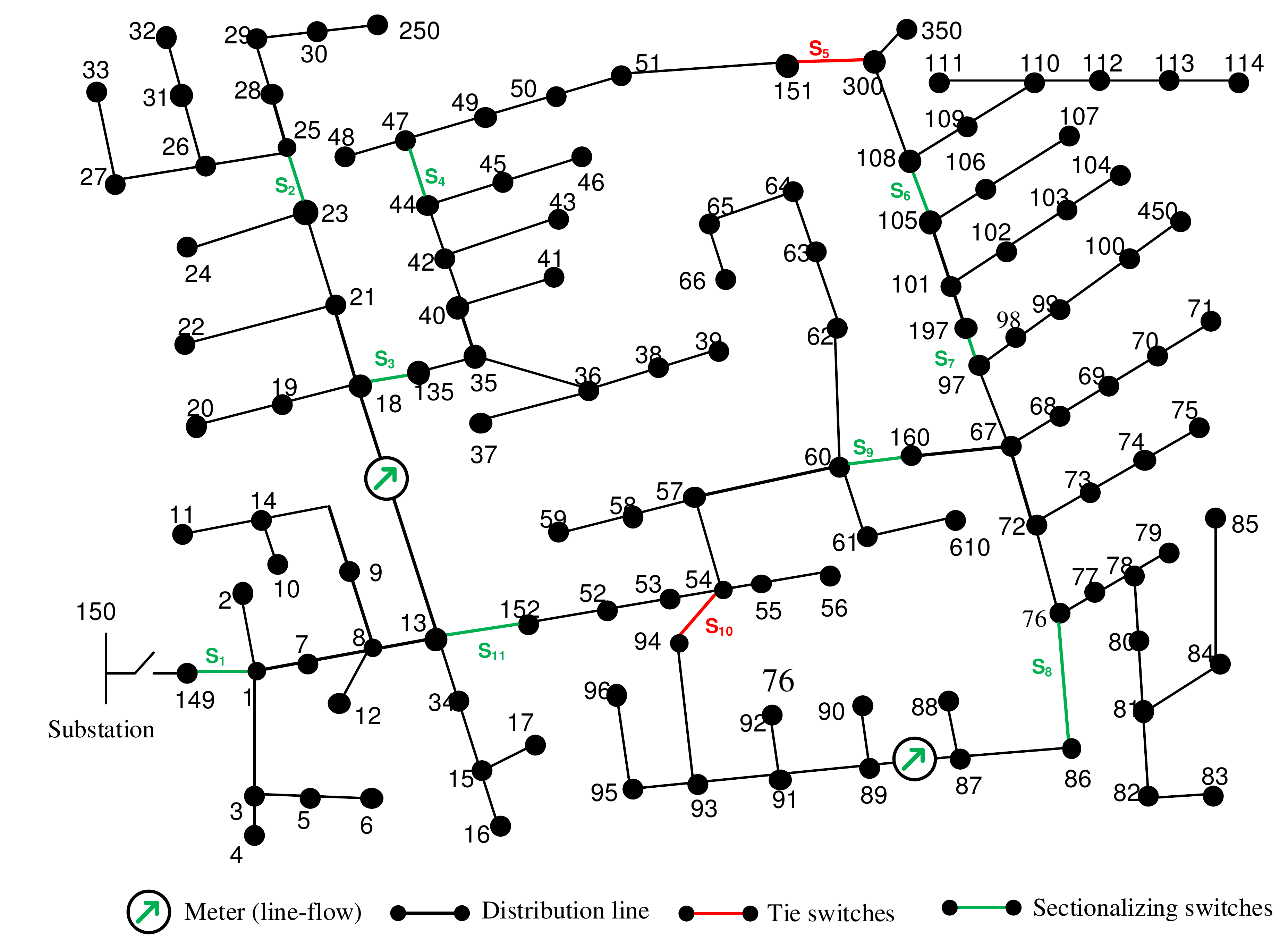}
	\vspace{-0.3cm}
	\caption{IEEE 123-bus feeder with measurement placement and showing location of switches}
	\vspace{-0.9cm}
	\label{fig:3}
\end{figure}
	
\vspace{-0.3cm}
	
\section{Results and Discussion}
topology estimation problem is formulated as an MILP which can be efficiently solved by using off-the-shelf solvers. We have used MATLAB R2018a to build the MILP model that is solved using inbuilt $intlinprog$ solver. The simulation is carried out on a PC with 3.4 GHz CPU and 16 GB RAM. In this study, modified IEEE 123-bus feeder \cite{ieee123} and multi-feeder test system consisting of four taxonomy feeder R3-12.47-2 \cite{multi_fed} are used to validate the proposed algorithm.  

\vspace{-0.4cm}
\subsection{Metrics}
The accuracy of the proposed approach is thoroughly tested for different percentages of measurement errors during both normal and outage conditions. The following three metrics are used to quantify the accuracy of the proposed topology estimation algorithm.
\begin{itemize}[leftmargin=*]
    \item{{\em Missed Detection Rate (MDR)}:  Out of $N$ possible topologies, if $N_{nc}$ is the number of topologies that are incorrectly estimated, then
    $\%MDR=\left(\frac{N_{nc}}{N}\times 100 \right)$. The incorrect estimation of a single switch status renders the given topology incorrectly identified.} 
    \item{{\em Mean Missed Switches (MMS)}: MMS measures the performance of the algorithm in correctly estimating switch statuses. Out of $N$ possible topologies, if $S_i$ switches are identified incorrectly then $\%MMS=\left(\frac{S_i}{S\times N} \times 100 \right)$, where, $S$ is the total number of switches.}
    \item{{\em Mean Missed Outages (MMO)}: MMO measures the performance of the algorithm in correctly detecting all outaged sections. Out of $N$ possible topologies, if $L_i$ sections are incorrectly estimated then $\%MMO=\left(\frac{L_i}{L_o\times N} \times 100 \right)$, where, $L_o$ is the total number of outaged sections.}
\end{itemize}

\vspace{-0.4cm}

\subsection{Simulation Steps}
Given a large number of possible operational topologies, numerous simulations are performed to realistically measure the performance of the algorithm. 
The simulation steps are detailed here. First, we randomly generate a radial topology for the given distribution test system that may or may not have outages. For the given topology, we run power flow using OpenDSS to obtain the measurement set. Next, we generate noisy measurements by adding Gaussian errors with zero mean and pre-specified variances to the flow and load measurements and Bernoulli errors to the smart meter ping measurements with a pre-specified error probability. For the simulation purpose, we study the performance of the proposed approach considering 1\%, 10\% \& 20\% error in load measurements and 0\%, 2\%, \& 5\% error in smart meter ping measurement for different cases. We solve the proposed MILP model to estimate the operational topology using the erroneous measurements. The performance metrics are calculated. The process is repeated several times until statistically significant results are obtained. On an average, it takes less than 30 sec to estimate the operational topology for the selected 1096-bus three-phase test system and less than 2 sec to estimate the operational topology for the selected IEEE 123-bus test case.
	
\begin{figure}[b]
	\centering
	\vspace{-0.5 cm}
	\includegraphics[width=0.50\textwidth]{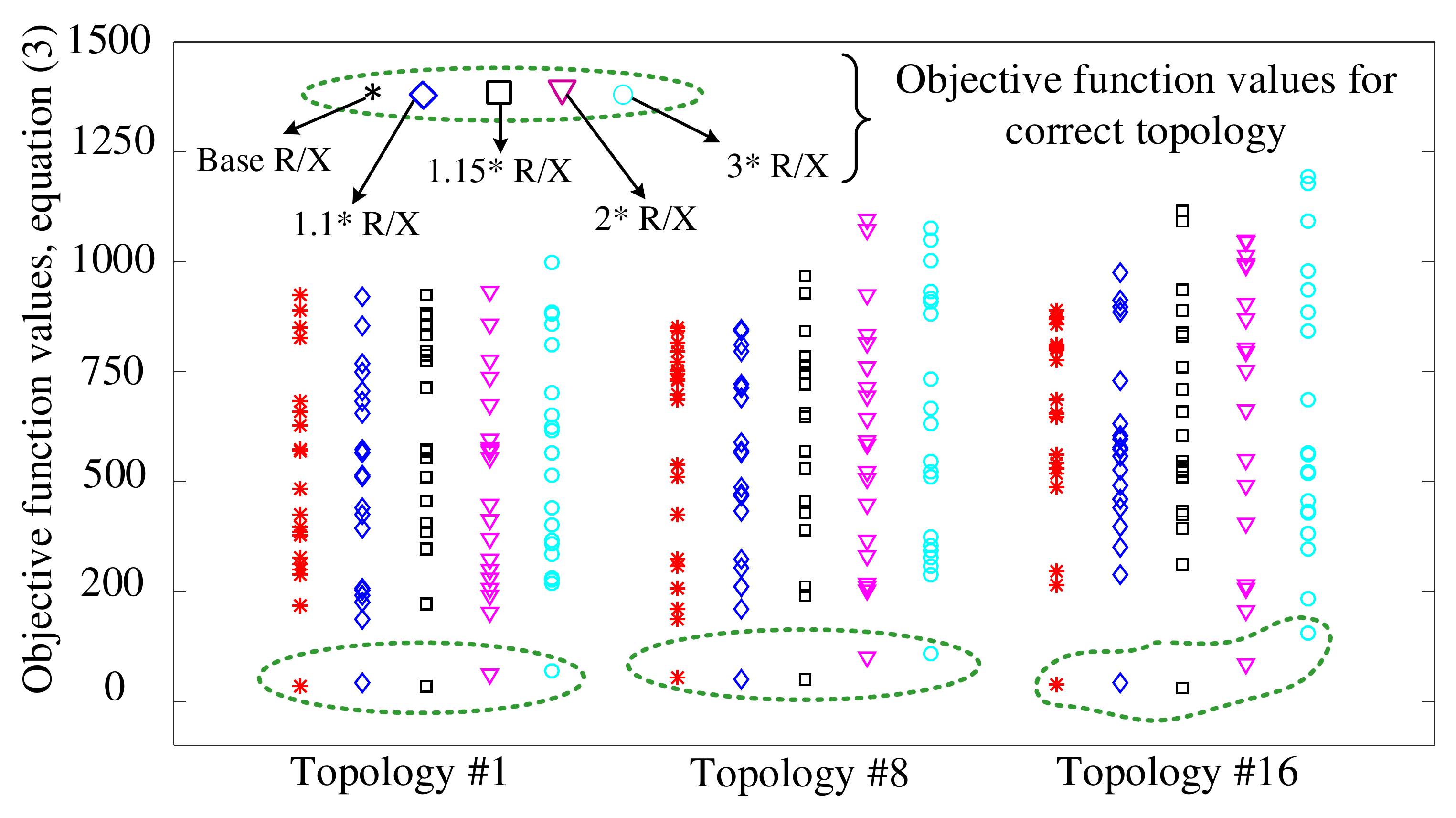} 
	\vspace{-0.8 cm}
	\caption{Influence of R/X parameters in topology identification. Even for increased R/X ratio, the correct topology and set of incorrect topology are clearly distinguishable.}
	\label{fig:loss}
	\vspace{-0.0 cm}
\end{figure}

\vspace{-0.4cm}

\subsection{Modified IEEE 123-bus Feeder}
IEEE-123 bus system is three-phase unbalanced distribution system model supplying multiple single and three-phase loads. The planning model for the original IEEE 123-bus system is modified by adding five new sectionalizing switches to create several possible operating topologies. Among the switches, $s_5$ and $s_{10}$ are normally open and rest are normally closed (See Fig. \ref{fig:3}). This system has 3 cycles and 20 possible normal radial operational topologies. To ensure topological observability, one power flow measurement unit is deployed in each fundamental cycle \cite{sevlian2015distribution}. Measurement locations in a cycle are randomly selected and shown in Fig. \ref{fig:3}.  
\begin{figure*}[t]
	\centering
	\includegraphics[width=0.99\textwidth]{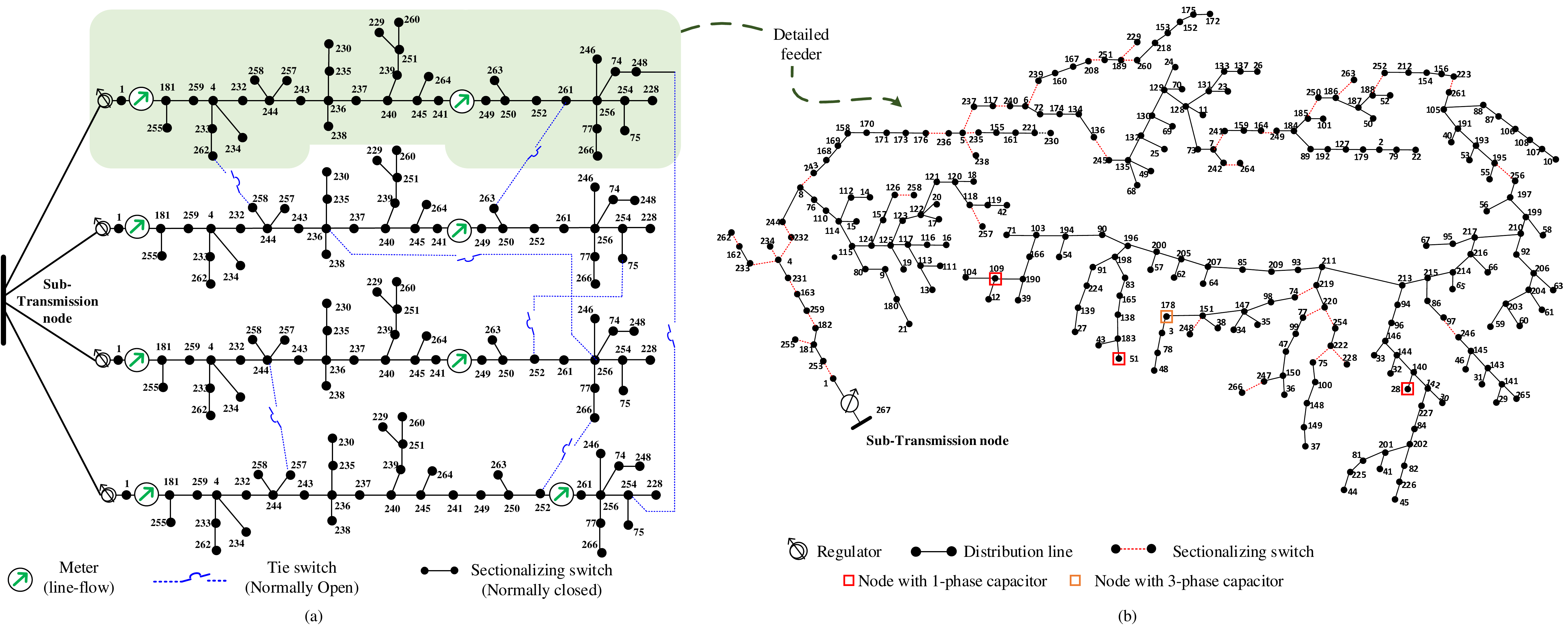} 
	\vspace{-0.5 cm}
	\caption{Test system: (a) Simplified model for multi-feeder 1069 bus test case with only switches in display to represent reduced diagram of each feeder, (b) Detailed R3-12.47-2 feeder model with distribution lines (black solid) and sectionalizing switches (red dashed) } 
	\label{fig:4}
	\vspace{-0.6 cm}
\end{figure*}

\subsubsection{Topology Estimation During Normal Operation}
During the normal operation, all topologies are correctly detected for the pre-specified error probabilities, i.e., for load measurement errors of 1\%, 10\% and 20\% and ping measurements errors of 2\% and 5\%. Note that the formulation discussed above approximates the power flow equations in the topology constrained power flow formulation. Therefore, we study the performance of the proposed approach for varying R/X ratio for distribution lines to evaluate the impact of power losses on the accuracy of the proposed topology estimation algorithm. To simulate these cases, we increase the R/X parameter for the distribution lines and observe whether the correct operational topology is identified or not. Also, we study how the change in R/X value affects the objective function value used in the topology estimation problem.  Note that the correct topology should result in the minimum value for the defined objective function that measures the difference between the measured and calculated power flow and load demand variables. Also, note that on increasing the R/X ratio, the losses in the network increases as shown in Table \ref{table:loss}; the percentage losses are calculated as the fraction of the total substation power demand.   
   
The proposed topology estimation algorithm is tested for all simulated test cases with varying R/X ratio shown in Table I.  It is observed that all topologies are still correctly identified on increasing the losses in the distribution system. We further elaborate on the observations using three randomly selected topologies. In this discussion, for a specific unknown topology, we calculate and plot the objective function value (i.e. minimization of errors, equation (\ref{eq:obj})) for all possible topologies. Fig. \ref{fig:loss} shows the objective function values all possible topologies for the three selected unknown topologies. Note that the correct topology leads to a minimum objective function value. The topology estimation is, therefore, able to identify the correct topology if the objective function value for the correct topology does not overlap with the objective function value for any of the incorrect topologies. As can be seen in Fig. 4, the objective function value for all incorrect topologies are much larger than for the correct topology. Therefore, we can identify the correct topology.

 \begin{table}[t]
        \centering
        \caption{Percentage loss for different R/X ratio. The average R/X ratio for base case is 0.4343.}
        \vspace{-0.2cm}
        \label{table:loss}
        \begin{tabular}{cccccc}
        \hline
            \toprule[0.3 mm]
            R/X ratio $\to$& Base&1.1$\times$R/X&1.15$\times$R/X&2$\times$R/X&3$\times$R/X\\
            \% Loss$\to$&4.25&4.92&5.08&7.47&11.37\\
            \hline
            \toprule[0.3 mm]
        \end{tabular}
        \vspace{-1.4cm}
\end{table}

Next, to study the effect of increasing R/X ratio, we again plot the objective function values for all topologies for the three selected cases. The objective function values for different R/X ratios are shown as parallel plots in Fig. \ref{fig:loss}. It is observed that even for increased R/X ratio, the objective function value for the correct topology is less than the values for incorrect topologies making it easily distinguishable from the rest of the topologies. Even if R/X ratio is increased by 200\%, the topologies are still distinguishable as the objective function value for correct and incorrect topologies do not overlap. Although the objective function values for correct and incorrect topologies move close to each other on increasing the losses, there is still a big margin and losses do not make the correct topology indistinguishable from incorrect topologies.
   
\begin{table}[t]
 	\caption{\%MDR, \%MMS and \%MMO for tested topologies with outages IEEE for 123-bus test system}
 	\vspace{-0.3cm}
 	\label{table:out}
 	\begin{tabular}{p{0.7cm}|p{0.45cm}p{0.45cm}p{0.45cm}|p{0.45cm}p{0.45cm}p{0.45cm}|p{0.45cm}p{0.45cm}p{0.45cm}}
 	    \toprule[0.03 cm]
 		\hline
 		\multirow{3}{0.9 cm}{\% error in $\hat{y}_l^\phi$}&\multicolumn{9}{c}{\% error in load measurements}\\
 		\cline{2-10}
  		&\multicolumn{3}{c}{\%MDR}&  \multicolumn{3}{c}{\%MMS} &\multicolumn{3}{c}{\%MMO}\\
	    \cline{2-10}
 		&1\%&10\%&20\%&1\%&10\%&20\%&1\%&10\%&20\%\\
 		\hline
 		0& 0&0&0&0&0&0&-&-&-\\
 		2& 0.047&0.143&0.238&0.004&0.013&0.022&0.015&0.015&0.015\\
  		5& 0.095&0.190&0.333&0.008&0.017&0.03&0.015&0.015&0.015\\
 		\toprule[0.03 cm]
 		\hline 
 	\end{tabular}
 	\vspace{-1cm}
 \end{table}
 
\subsubsection{Topology Estimation During Outage Condition}
For each topology, one fault is randomly simulated and the algorithm is repeatedly tested 10 times for each scenario.  Table \ref{table:out} shows the performance of the approach in the presence of errors in load and flow measurements. For 1\% error in load measurements and no error in ping measurements, \%MDR is 0 indicating that there are no misdetected topologies and all switches and outages are accurately identified. While for the worst case of 20\% error in load measurements and 5\% error in smart meter measurements, \%MDR is 0.333 implying that out of 2100 tested scenarios, 7 topologies are misdetected. This amounts to 0.03\% misdetection of switches (\%MMS) and  0.015\% of missed outages (\%MMO).

One particular scenario for misdetection is detailed here. Let us consider the case where the operational topology (sub-tree) has switches  $s_2$, $s_6$, and $s_{10}$ open with other switches closed. However, the optimization detected the correct operation topology having switch $s_5$ open instead of $s_2$. As observed, the algorithm detected $s_2$ as closed and supplying the load section downstream when it shouldn't and meanwhile switch $s_5$ is detected as open while it's actual status is closed and is supplying the section of load between $s_6$ and $s_5$. To further elaborate the reason for this misdetection, we observed the sectional loads that are supplied downstream from $s_2$ and between switches $s_5$ and $s_6$. It is observed that both sections are supplying for the approximately equal loads and therefore, the meter on line 13-18 (which supplies for both load sections) is unable to distinguish the correct section as it reads almost the same power flow for both topologies. Therefore, for the cases when flow meters cannot distinguish the load sections (due to similar load demands), the proposed approach detects an incorrect operational topology.

 \vspace{-0.3cm}
 
\subsection{Multi-Feeder Test System}
The proposed topology estimation algorithm is also demonstrated using a three-phase unbalanced 1069-bus multi-feeder test system. Four taxonomy feeders R3-12.47-2 \cite{multi_fed} are replicated to obtain the four-feeder 1069-bus distribution system connected using seven normally open tie switches (See Fig. \ref{fig:4}a). Detailed model of one of the four feeders is shown in Fig. \ref{fig:4}b. Each feeder has 40 three-phase sectionalizing switches (normally-closed switches). Seven tie-switches and 40 sectionalizing switches lead to a large system with the number of operational topologies in the order of millions.  
Each feeder has three single-phase capacitors and one three-phase capacitor at locations shown in Fig. \ref{fig:4}. The operational statuses of capacitor banks are unknown when solving the topology estimation problem. 
The measurement set is described next. Topological observability requires one flow measurement per fundamental simple cycle. A randomly selected measurement placement satisfying this criterion is generated as shown in Fig. \ref{fig:4} and is used for testing. A total of 10\% of smart meters are pinged from each load section. This measurement placement is tested with 1\%, 10\% and 20\% errors in load measurements and for 0\%, 2\% and 5\% errors in ping measurements. 

\begin{figure}[t]
	\centering
	\vspace{-0.0cm}
	\includegraphics[width=0.45\textwidth]{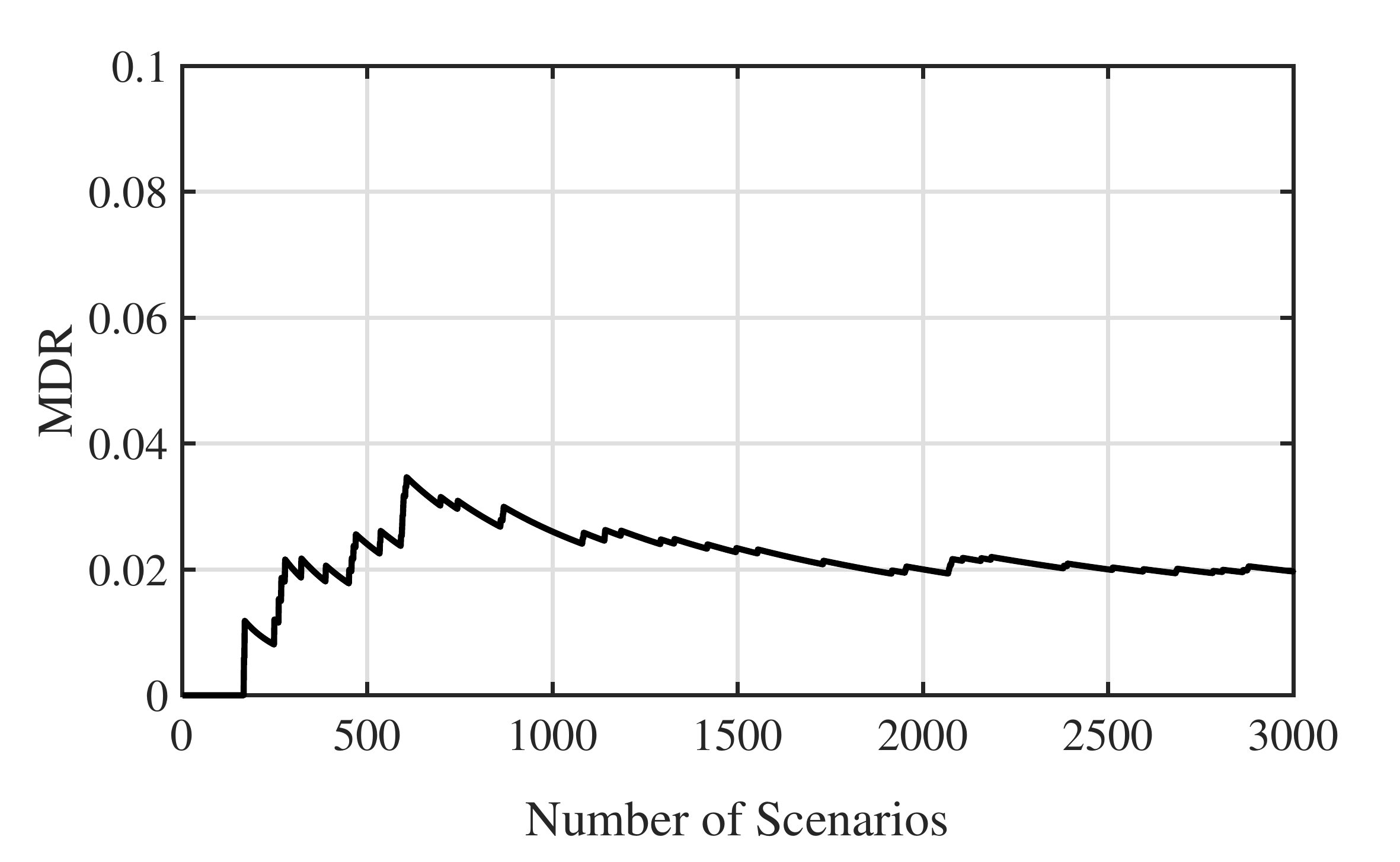} 
	\vspace{-0.5cm}
	\caption{Plot showing convergence of MDR with topologies tested}
	\vspace{-1cm}
	\label{fig:5}
\end{figure}
	 
\subsubsection{Topology Estimation During Normal Operation}
In this section, the proposed algorithm is tested for its accuracy in detecting operational topology during normal operation (i.e., without any outages). For an exhaustive validation of the algorithm, a total of 3000 test scenarios are simulated by randomly sampling an operational topology as detailed in Section IV.A. To ensure that test set-up represents statistically significant test scenarios, a plot for the number of topologies tested vs. MDR for a scenario with worst-case error (20\% error in load measurement and 5\% error in ping measurements) is plotted (see Fig. \ref{fig:5}). Note that MDR is almost constant for the number of tested scenarios greater than 2500 which justifies our choice of testing for 3000 scenarios.
	 
\begin{table}[!t]
 \centering
 \caption{\%MDR and \%MMS for tested topologies without outages for 1096-bus test system}
 \vspace{-0.3cm}
 \label{table:2}
 \begin{tabular}{p{0.7cm}|p{0.8cm}p{0.8cm}|p{0.8cm}p{0.8cm}|p{0.8cm}p{0.8cm}}
 \toprule[0.03 cm]
 	\hline
 	\multirow{3}{0.9 cm}{\% error in $\hat{y}_l^\phi$}&\multicolumn{6}{c}{\% error in load measurements}\\
 	\cline{2-7}
 	&\multicolumn{2}{c|}{1\%}&\multicolumn{2}{c|}{10\%}&\multicolumn{2}{c}{20\%}\\	
 	\cline{2-7}
 	&\%MDR&\%MMS&\%MDR&\%MMS&\%MDR&\%MMS\\
 	\hline	
 	0&0&0&0&0&0.5&0.0182\\
 	2&0&0&0.0311&0.0011&1.50&0.0737\\
 	5&0&0&0.0667&0.0024&1.53&0.0752\\
 	\toprule[0.03 cm]
 	\hline
 \end{tabular}
\vspace{-1.5cm}
\end{table}

From results in Table \ref{table:2}, we see that as expected there is no missed detection when there is only 1\% error in load measurements and no errors in smart meters. Note that the misdetection is still zero when the error in smart meters is increased to 10\% while keeping the error in load measurement to 1\%. The MDR increases on increasing the error in load measurements while keeping the errors in ping measurements as zero. For 10\% error in load measurements, all the topologies are still distinguishable but when the errors increased to 20\%, few topologies are incorrectly detected. 
On further investigation, it is observed that at most one switch pair is incorrectly identified for all topologies that are misdetected. Thus, MMS for all cases is significantly low as compared to MDR (see Table \ref{table:2}). 
	  
Next, the error in ping measurements is also increased to 5\% and 10\%. It is observed from Table \ref{table:2} that the \%MDR is slightly increased on increasing the errors in ping measurement. The \%MMS still remains small indicating that misdetection is largely due to the incorrect status of a small number of switch pairs. Based on small \%MMS, we can conclude that the proposed algorithm is highly accurate in estimating the switch statuses even with high-levels of measurement errors.

\subsubsection{Topology Estimation During Outage Condition}
We randomly select a possible normal topology and simulate any random number of faults between 1 to 3 at random locations; 3000 such random outaged topologies are simulated. For each outaged topology, the topology estimation algorithm is solved for different percentages of errors in load and smart meter measurements. The results are tabulated in Table \ref{table:3}.

As expected, for 1\% error in load measurements and no error in smart meter measurements, all outaged topologies are detected. Unlike normal operation, the misdetection is no longer zero when the error in smart meters is increased while the error in load measurements is still at 1\%. Next, the error in load measurements is increased while keeping smart meters error-free. For 10\% error in load measurements, all the topologies are still distinguishable but when the error is increased to 20\%, few topologies are incorrectly detected (see Table \ref{table:3}). 

\begin{table}[t]
 \caption{\%MDR, \%MMS and \%MMO for tested topologies with outages for 1096-bus test system}
 \vspace{-0.3cm}
 \label{table:3}
 \begin{tabular}{p{0.7cm}|p{0.4cm}p{0.4cm}p{0.4cm}|p{0.4cm}p{0.4cm}p{0.4cm}|p{0.4cm}p{0.4cm}p{0.4cm}}
 \toprule[0.03 cm]
 	\hline
 	\multirow{3}{0.9 cm}{\% error in $\hat{y}_l^\phi$}&\multicolumn{9}{c}{\% error in load measurements}\\
 	\cline{2-10}
  	&\multicolumn{3}{c}{\%MDR}&  \multicolumn{3}{c}{\%MMS} &\multicolumn{3}{c}{\%MMO}\\
\cline{2-10}
 	&1\%&10\%&20\%&1\%&10\%&20\%&1\%&10\%&20\%\\
 	\hline
 	0& 0&0&1.03&0&0&0.04&0&0&0.07\\
 	2&0.53&2.80&7.10&0.019&0.11&0.35&0.041&0.32&0.73\\
  	5&0.91&4.35&7.23&0.041&0.17&0.37&0.067&0.40&0.78\\
 	\toprule[0.03 cm]
 	\hline 
 \end{tabular}
 \vspace{-0.5cm}
\end{table}
     
It can be observed from Table \ref{table:3} that the \%MDR during outages increases significantly with the increase in ping measurement errors. Note that MDR is an extremely conservative metric. For any topology, even if the single switch is incorrectly identified it is counted as a misdetected topology in MDR calculation. Therefore, we elaborate the strength of the algorithm in its capability to correctly estimate outaged sections and switch statuses using the other two metrics: MMS and MMO. 
     
\begin{figure}[!t]
	\centering
	\vspace{0.1 cm}
	\includegraphics[width=0.45\textwidth]{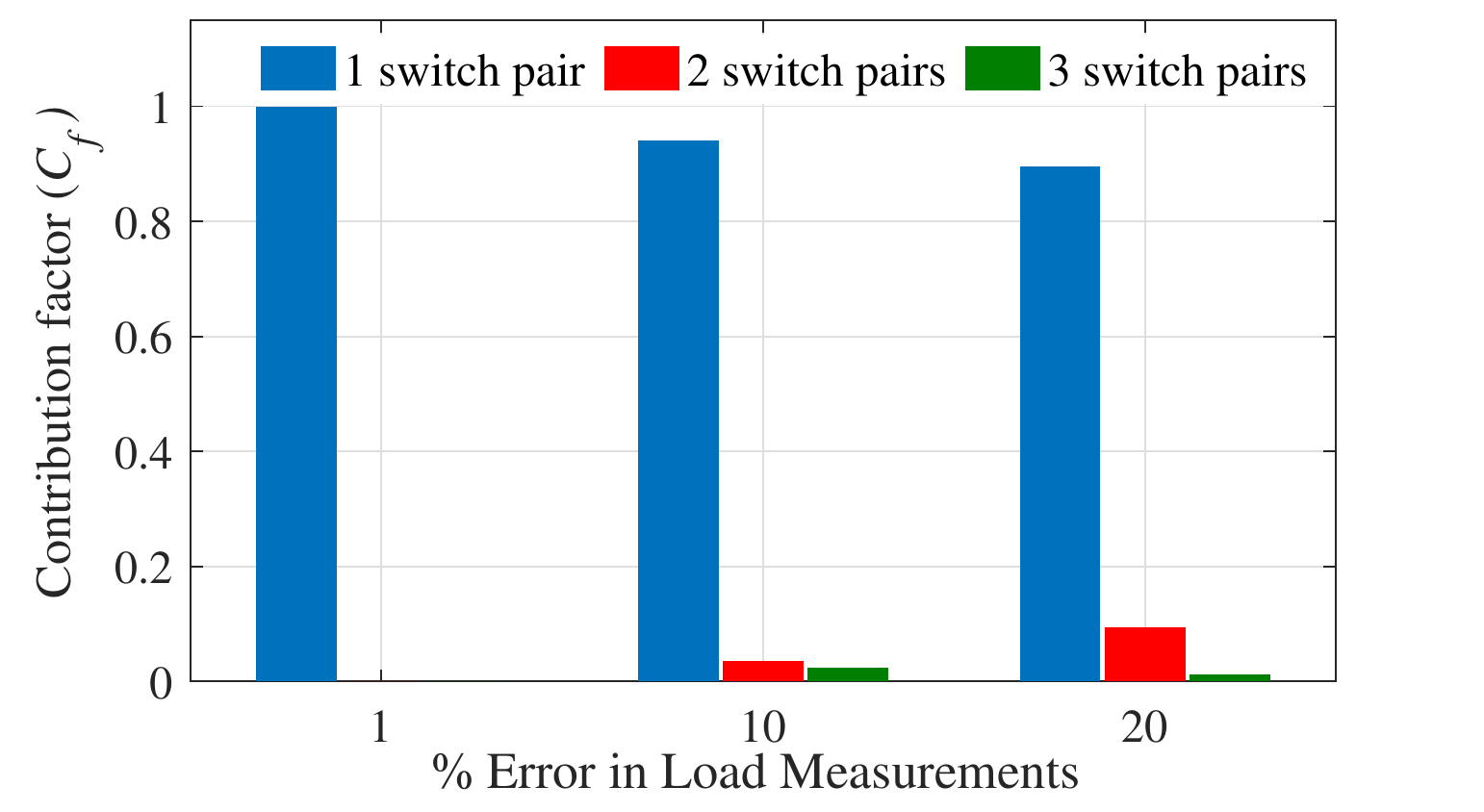} 
	\vspace{-0.4 cm}
	\caption{Contribution to \%MDR from different number of switch pairs.}
	\label{fig:6}
	\vspace{-0.4cm}
\end{figure}

From Table \ref{table:3}, it can be noted that the \%MMO is significantly low even for worst-case measurement errors in loads and pings implying that the algorithm can correctly estimate most of the outaged sections. Also, MMS is low indicating most topologies are misdetected by very few numbers of incorrect switch statuses. To further elaborate the result, we observe the number of switch pairs that are incorrectly estimated for each misdetected topology.
For 5\% error in smart meter measurement, out of the total misdetections, contribution factor ($C_f$) of misdetected switch pairs for different errors in load measurements is shown in Fig. \ref{fig:6}. Note that $C_f$ is the fraction of misdetection contributed by different switch pairs to the total misdetection. It can be observed that most of the misdetections corresponded to only one pair of switches being incorrectly estimated. The proposed algorithm is, therefore, able to correctly estimate most of the switches and outaged sections even with high-levels of measurement errors.  

\begin{figure}[!t]
    \centering
    \includegraphics[width=0.45\textwidth]{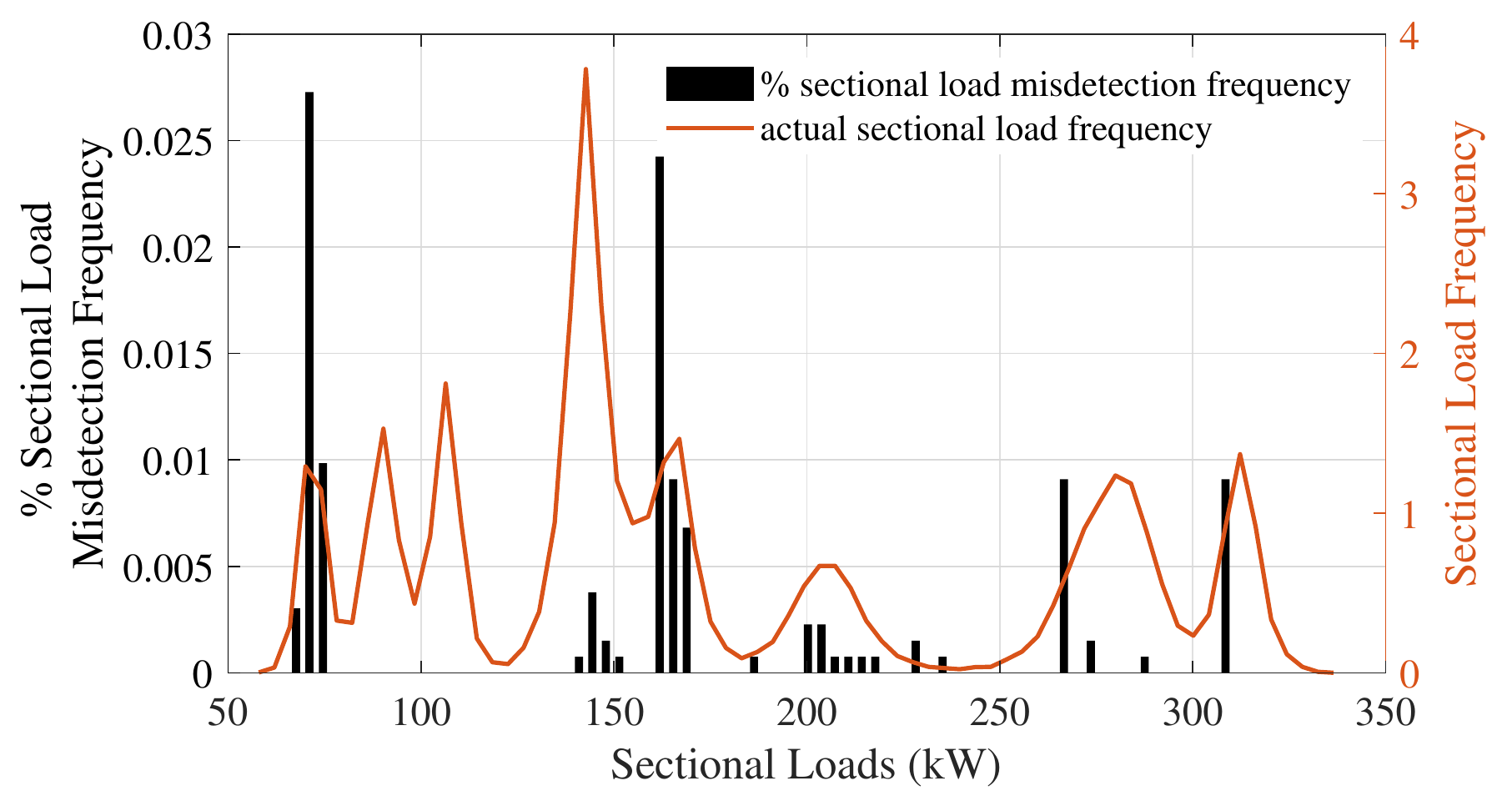} 
    \vspace{-0.4 cm}
    \caption{Sectional load misdetection: 10\% load measurement error.}
    \label{fig:7}
    \vspace{-1.5cm}
\end{figure}

We further explore the reason for misdetection during an outage by identifying the load sections that are most frequently misdetected. We observe that the errors in smart meter measurements make some of the healthy and outaged sections with similar load demand indistinguishable. This is because the flow measurement in the upstream line is nearly the same for an outage in either load. This observation is elaborated using the case with 10\% error in load measurements and 5\% error in ping measurement. A histogram for the sectional loads with actual and erroneous data is plotted in Fig. \ref{fig:7}. The histogram represents the frequency of observing a load section of given kW demand. 
Next, we calculate the percentage of times each sectional load is misdetected (bar plot in Fig. \ref{fig:7}). 
From Fig. \ref{fig:7} it can be observed that the most frequently misdetected sections are indeed the ones that are most frequently observed; the bar plot overlaps with the spikes of a histogram for sectional loads. Note that there are a few sections that are never misdetected even though they are frequently observed. This is because they are never supplied by the same upstream flow meter.

 \vspace{-0.2cm}
 
\section{Conclusion}
The existing literature on topology estimation cannot simultaneously estimate the distribution grid’s operation topology and outage sections in a computationally tractable manner. This paper presents a generalized algorithm to estimate the operational topology during both normal and outage conditions. The problem is formulated as an MILP to minimize the weighted error between the measurements and the system variables subject to topology constrained three-phase linearized power flow equations. The method relies on the pseudo load measurements, one power flow measurement on each cycle, and at least one smart meter measurement from each load section. The algorithm is thoroughly tested for a large-scale 1069-bus multi-feeder three-phase unbalanced test system for different percentages of measurement errors. For the test system, an operational topology is obtained on an average within 30 sec validating the applicability of the approach as a real-time topology processor. Furthermore, the results validate that the proposed approach is sufficiently accurate even with high percentages of measurement errors. 

\bibliographystyle{ieeetr}
\bibliography{references}

\vspace{-2.5 cm}
	
\begin{IEEEbiography}
[{\includegraphics[width=1in,height=1.25in,clip,keepaspectratio]{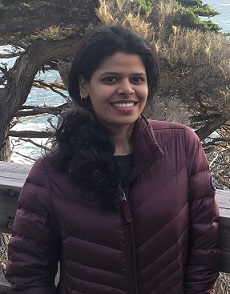}}]
{\textbf{Anandini Gandluru}} (S'16) 
received the B.Tech. degree in Electrical Engineering from G. Narayanamma Institute of Technology and Science, Hyderabad, India in 2013, and the M.Tech. degree from the Department of Electrical Engineering, National Institute of Technology, Warangal, Telangana, India. She received the M.S. degree from the School of Electrical and Computer Science, Washington State University, Pullman, WA, USA. Her research interests include state estimation and distribution system topology estimation. 
\end{IEEEbiography}

\vspace{-2.5 cm}

\begin{IEEEbiography}[{\includegraphics[width=1in,height=1.25in,clip,keepaspectratio]{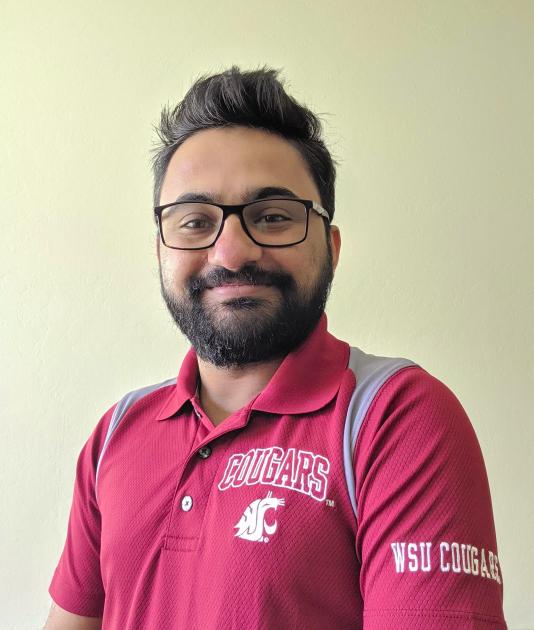}}]{\textbf{Shiva Poudel}} (S'15) received the B.E. degree from the Department of Electrical Engineering, Pulchowk Campus, Kathmandu, Nepal, in 2013, and the M.S. degree from the Electrical Engineering and Computer Science Department, South Dakota State University, Brookings, SD, USA, in 2016. He is now pursuing the Ph.D. degree in the School of Electrical Engineering and Computer Science, Washington State University, Pullman, WA. 
In 2018 and 2019, he was a summer intern with Mitsubishi Electric Research Laboratories, Cambridge, MA, USA and Electric Power Research Institute, Palo Alto, CA, USA respectively.
His current research interests include distribution system restoration, resilience assessment, and distributed algorithms.
\end{IEEEbiography}

\vspace{-2.5 cm}

\begin{IEEEbiography}[{\includegraphics[width=1in,height=1.25in,clip,keepaspectratio]{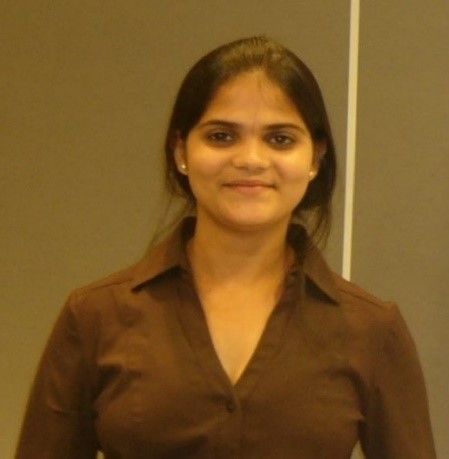}}]{\textbf{Anamika Dubey}} (M'16) received the M.S.E and Ph.D. degrees in Electrical and Computer Engineering from the University of Texas at Austin in 2012 and 2015, respectively. Currently, she is an Assistant Professor in the School of Electrical Engineering and Computer Science at Washington State University, Pullman.

Her research focus is on the analysis, operation, and planning of the modern power distribution systems for enhanced service quality and grid resilience. At WSU, her lab focuses on developing new planning and operational tools for the current and future power distribution systems that help in effective integration of distributed energy resources and responsive loads. 
\end{IEEEbiography}

\end{document}